\documentclass{aa}
\usepackage{threeparttable}
\usepackage{amssymb}
\usepackage{graphicx}
\usepackage{natbib}

\bibpunct{(}{)}{;}{a}{}{,}  

\newcommand{\eqb}{\begin{eqnarray}}
\newcommand{\eqe}{\end{eqnarray}}
\newcommand{\sth}{\sigma_{\rm T}}

\newcommand{\gemx}{\gamma_{\rm e,max}}

\newcommand{\mpr}{m_{\rm p}}

\newcommand{\tcr}{t_{\rm cr}}

\newcommand{\pks}{ PKS~2155-304}

\title{Time-dependent modelling of PKS~2155-304 in a low
state}
\author{M. Petropoulou$^{1,2}$}
\institute{$^1$ Department of Physics, Purdue University, 525 Northwestern Avenue, West Lafayette, IN, 47907, USA\\
$^2$NASA Einstein Postdoctoral Fellow
}
\date{Received.../Accepted...}

\abstract{}
{We apply both leptonic and leptohadronic emission scenarios for modelling the multiwavelength photon spectra
and the observed variability in the optical,
X-ray, and TeV gamma-ray energy bands of blazar PKS~2155-304 while being in a low state  between 25 August and  6 September 2008.
}
{We consider three emission models, namely a one-component synchrotron self-Compton  model (1-SSC),
a one-zone proton synchrotron model (LHs), and a two-component SSC model  (2-SSC).
Only in the first scenario can the emission from the optical up to very high-energy (VHE) gamma-rays
be attributed to a single particle population from one emission region. In the LHs model, the low-energy and high-energy
bumps of the spectral energy distribution (SED) are the result of electron and proton synchrotron radiation, respectively,
i.e. two different particle populations are required. In the 2-SSC model, the emission from one component dominates in the optical
and gamma-ray energy bands, while the other one contributes only to the X-ray flux. Using a time-dependent numerical code
that solves the kinetic equations for each particle species, we derived, in all cases, acceptable fits
to the time-averaged SED. By imposing variations to one (or more) model parameters according to observed variability pattern 
in one (or more) frequencies we calculated the respective lightcurves and compared them with the observations.
}
{We show that the 1-SSC model cannot account for the anticorrelation observed between the X-rays and VHE gamma-rays, although
it can explain the time-averaged SED. The anticorrelation can be more naturally explained 
by the two-component emission models. Both of them reproduce satisfactorily the optical, X-ray, and TeV variability but at the cost
of additional free parameters,  which from four in the 2-SSC model increase to six in the LHs model. 
Although  the results of our time-resolved
analysis do not favour one of the aforementioned models, they suggest that a two-component
scenario is more adequate for the emission of PKS~2155-304 in the low state of 2008, which agrees with 
a recent independent analysis. This suggests that the quiescent blazar radiation might result
from a superposition of the radiation from different
components, while a flare might still be the result of a single
component.
}
{}

\titlerunning{Time dependent modelling of PKS~2155-304 in a low state}
\authorrunning{M. Petropoulou} 

\begin{document}

\maketitle

\section{Introduction}
Blazars are a subclass of active galactic nuclei 
with a non-thermal emission covering most
of the electromagnetic spectrum, i.e. from radio up
to very high-energy (VHE) gamma-rays, and the 
dominant class of extragalactic sources at energies
$> 100$~MeV \citep{hinton09, holder13}.
Their broadband emission,
which originates from a
relativistic jet oriented close to 
our line of sight, 
is Doppler boosted and so it shows
no evidence of spectral lines, at least for the subclass of BL~Lac objects.
The spectral energy distribution
(SED) of TeV-emitting blazars consists of two smooth,
broad components (e.g. \citealt{ulrichetal97, fossatietal98}). The first one extends from
the radio up to the X-rays with a peak between
the optical  and soft X-ray energy bands, while the second one
extends up to TeV energies, with a peak energy around
$0.1$ TeV, although this is not always clear \citep{abdoetal11}.

An important tool in our attempt to understand the physics
of blazar emission is the modelling of their SED. 
Although it is a common belief that the lower energy bump is the
synchrotron emission 
of relativistic electrons, the origin of the
high-energy component is still under debate. 
Theoretical models are divided into leptonic and
leptohadronic, according to the type of particles responsible
for the gamma-ray emission. In  leptonic scenarios, the high-energy component 
is the result of  inverse Compton scattering of electrons
in a photon field. 
As seed photons can serve the synchrotron photons produced by the same electron population
(SSC models; e.g. \citealt{maraschietal92, konopelko03}) or/and photons  
from an external region (EC models), for example from the accretion disk 
\citep{dermeretal92,dermerschlickeiser93} or from the broad line
region \citep{sikoraetal94, ghisellinimadau96, boettcherdermer98}. 
In leptohadronic scenarios, on the other hand, 
the gamma-ray emission can be the result of (i) 
proton synchrotron radiation \citep{mannheimbiermann92, aharonian00, mueckeetal03}; (ii) 
synchrotron radiation of secondary pairs produced in the hadronic cascade (e.g. \citealt{petromast12, mastetal13});
or even of (iii) neutral pion decay \citep{sahu13,gao14}. For a recent review on leptohadronic modelling, see \cite{boettcher12, boettcher13}.

Stationary one-zone models have been extensively used  and, in most cases, both leptonic and hadronic models prove equally
successfull in fitting the SED of blazars 
\citep{maraschietal92, sikoraetal94, bloommarscher96, tavecchio98, cerruti12, 
reimer12, boettcher13, dimitrakoudis14, gao14}.
However, the recently obtained wealth of (quasi)simultaneous
observations that cover both the low-and high-energy regimes of the SED,
gives an opportunity for time-dependent modelling. Since leptonic and hadronic models are expected to have different
time signatures, this time-resolved fitting analysis can pose
new challenges to both categories of models. It can be used either to explain
the origin of a flare (see e.g. \citealt{krawczynski02} and references therein)
or to lift the degeneracy between models that can successfully fit the time-averaged SED \citep{mastetal13},
yet, time-dependent blazar modelling has not been widely applied to
quiescent emission mainly because of the lack of contemporeneous MW data in low states.

In the present work we aim to test whether one-zone emission models can account for both
the observed SED and lightcurves of 
a blazar source, even when this is in a low state where the variability is marginal.
We investigate, in particular, three possible scenarios in a time-dependent way: (i)
a one-zone SSC model, where the emission from the optical up to TeV gamma-rays is explained
in terms of a relativistic electron distribution; (ii) a one-zone leptohadronic model (LHs), where
the low-energy and high-energy humps are the result of electron and proton synchrotron radiation, respectively; and (iii)
a two-component SSC model, where we assume that there are two physically distinct regions that contribute to the overall SED.

We apply our models to the multiwavelength observations
of the high-peaked blazar (HBL) \pks \ at redshift $z=0.116$ in a low state \citep{aharonian09} -- henceforth, A09. 
The particular choice of data was motivated by the following: (i) the blazar was for the first time 
monitored simultaneously in four energy bands, namely in optical with ATOM \citep{hauser04}, in X-rays with RXTE \citep{jahoda96} and {\sl Swift} \citep{burrows05}, 
in GeV gamma-rays with {\sl Fermi} \citep{atwood09} and in TeV gamma-rays with H.E.S.S. (A09); (ii)
it was observed in a low state with marginal variability at least
at two energy bands (optical and GeV gamma-rays) implying that the underlying physical
conditions do not vary significantly; (iii) a significant correlation between
the optical and VHE gamma-ray fluxes was found, which is not commonly observed in HBLs \citep{krawczynski04, 
wagner08, aharonian09b} -- see, however, \citealt{donnarumma09}
for a possible correlation observed during a flare of Mrk~421;  and (iv) no correlation between
the X-rays and VHE gamma-rays was detected in contrast to flaring events \citep{aharonian09b, abramowskietal12}. 
As we show in \S\ref{results}, it is the last two observational facts, in particular, that
will be used to distinguish between the one- and two-component emission models.

For each one of the models, we find first a set of parameters
that lead the system to a steady-state that fits in a broad sense the
time-averaged SED. This is used as an initial condition for studying the properties of
the source in the period MJD 54704-54715. For this, we vary
one or more model parameters according to the temporal pattern observed
in one or more energy bands and try to reproduce the observed variability in as many as possible energy bands. 
Although first results can be found in \cite{petromast13}, here we 
extend the previous analysis by discussing in more detail 
the emission models and by presenting fiducial flaring episodes in the context of the 2-SSC model.
We note also that \cite{almeida14} used the same dataset as an application of their
method, which is based on studying the properties of optical polarization from blazar sources \citep{almeida10},
and, interestingly, the authors reached to similar conclusions to ours.

The present work is structured as follows: in \S2 we describe the general framework and
present in more detail our method in \S3. We continue with the presentation of our results 
in \S4 and conclude in \S5 with a summary and discussion.
Throughout this study we use $H_0=70$~km Mpc$^{-1}$ s$^{-1}$, $\Omega_{\rm M}=0.3$ and 
$\Omega_{\Lambda}=0.7$.

\section{The model}
\label{model}
In what follows we use the one-zone leptohadronic model as described in 
\cite{DMPR12} -- henceforth DMPR12 and for completeness we repeat here its basic points.
We  consider a spherical blob of fixed radius $R$ moving with a Doppler factor
$\delta$ with respect to us and containing a magnetic field of strength $B$. 
We also assume that relativistic electrons and/or protons are
being injected in the blob with a rate given by
\eqb
Q_{i}(\gamma, \tau) = Q_0\gamma^{-s_i} e^{-\left(\frac{\gamma}{\gamma_{i,\max}}\right)^{q_i}} H(\gamma-\gamma_{i, \min})H(\tau), 
\label{qinj}
\eqe
where the subscript $i=p,e$ accounts for protons and electrons, $s_i$ is the power-law index,
$H(x)$ is the Heaviside function, $\tau$ is the comoving time measured in units of the crossing time
$\tcr=R/c$, and $Q_0$  is the amplitude of the injection rate.
By setting $Q_p=0$ the leptohadronic model simplifies into the usual SSC model\footnote{As there is no evidence of a strong
external photon field in the environment of blazar \pks, we use the terms ``leptonic'' and ``SSC'' models interchangeably.}.
The minimum and maximum Lorentz factors of the distribution are denoted as $\gamma_{i,\min}$ and $\gamma_{i,\max}$, respectively. 
The exponent $q_i$ determines the curvature of the exponential cutoff and typical values predicted by 
acceleration models are $1-2$ (see e.g.~\citealt{lefa11}).
Finally, particles are also allowed to leave the region after a characteristic timescale $t_{i, \rm esc}=\tcr$.

The injection rate can be associated with a compactness\footnote{Expression (\ref{linj}) contains a factor of three not present in the conventional
definitions (see e.g. \citealt{petromast11})} as
\eqb
\ell_{i}^{\rm inj}=\frac{3 L_{i}^{\rm inj} \sth} {4\pi R m_i c^3}, 
\label{linj}
\eqe
 where the injected luminosity in protons or electrons as measured in the comoving frame
 is given by
 \eqb
 L_{i}^{\rm inj}=V_b m_i c^2 \tcr^{-1} \int_{\gamma_{i,\min}}^{\infty} d\gamma \gamma Q_i(\gamma,\tau),
 \label{Luminj}
 \eqe
 where the factor $\tcr$ is introduced because the injection rate defined in Eq.~(\ref{qinj}) is expressed in terms of the dimensionless time $\tau$.

Protons can lose energy via three channels:
(i) synchrotron radiation; (ii) photopair production (Bethe-Heitler); and 
(iii) photopion production. The relative effect of these three processes on the proton distribution function 
depends on the specific parameters of the system, therefore all three have 
to be taken into account in a kinetic equation, which besides the proton injection term, contains a proton escape term. 
Leptohadronic modelling is far more complex compared to SSC modelling mainly because of the 
creation of secondary particles, such as pions, which eventually decay to electron/positron
pairs. Thus, one has to also follow the evolution of photons and electrons, by writing two additional kinetic equations for them that are being coupled
to the equation of protons. Neutrons and neutrinos are also byproducts of photopionic
interactions and, in principle, one has to include two more
equations for them -- see DMPR12. In this study, where the dominant energy loss process for protons
is synchrotron radiation, one can safely ignore them. 

At the cost of no spatial information, i.e. by assuming uniform distributions inside the blob,
one can obtain a detailed picture of the particle distribution evolution in time and in energy by
solving a system of coupled partial integrodifferential equations, which  for leptohadronic models consists at least
from five equations. However, in the case of SSC models, 
the system simplifies into a set of two equations, one
for photons and one for electron/positron pairs.

The above scheme 
can be used to derive both steady-state and time-dependent solutions.
If all parameters are constant in time the system will 
eventually reach a 
steady-state\footnote{This is partially true  for  leptohadronic models, since for particular parameter sets
the system can exhibit limit cycle behaviour
\citep{mastetal05, PM12}.}. If, on the other hand, 
we allow for one or more parameters to have some explicit time
dependence, such as $Q_0$ and $\gamma_{i,\max}$ in Eq.~(\ref{qinj}), 
then the system will not reach a steady state but it
will show temporal
variations, which will be associated with the variations imposed on the
input parameters.

\section{The method}

The method we follow is similar to that described in \cite{krawczynski02} in that
we attempt a time-dependent modelling of the observed lightcurves in one or more
energy bands. Instead of  approximating the time-variability by a sequence 
of steady states, we calculate
the evolution of the electron distribution and
the resulting photon spectra at each crossing time, which serves as a unit
for the timestep used in the numerical code (for more details, see \citealt{mastkirk95} and DMPR12). Moreover, the present work
extends the previous analysis of \cite{krawczynski02} by attempting time-dependent modelling of MW observations 
in the context of a leptohadronic model. It is noteworthy that for \pks, in particular, 
only stationary solutions for its MW spectra were obtained within the leptohadronic context \citep{cerruti12}. 

In order to model the MW data of \pks \ during the period  MJD 54704-54715, we have considered
the following three models listed in order of increasing
complexity: (i) one-component SSC model (1-SSC), (ii)  leptohadronic proton synchrotron model (LHs) and (iii)  two-component 
SSC model (2-SSC). In all cases we have employed a four-step process: 
\begin{enumerate}
\item  We determine a set of parameters constant in time that leads the system to a steady state that 
lies close to but below the time-averaged SED.
\item  We use this steady state as an initial condition of the system. 
 \item We impose variations to one or more parameters following the variability pattern observed in one or more energy bands.
 \item We calculate the lightcurves for different values of the parameters that are related to the imposed variations. In case
 we do not find an acceptable fit to at least one of the observed lightcurves we go to step (1) and repeat the procedure.
\end{enumerate}
We use the term ``acceptable fit'' to emphasize that
we did not search the whole parameter space for pinpointing the set giving the best $\chi^2$ fit. This would
not serve the purpose of this work, which can be summarized in the following: to 
calculate model spectra and lightcurves 
that can roughly account for both the time-averaged SED and the observed 
variability of \pks \ and use them as a stepping stone for a qualitative comparison between the models.
In any case, we verified that the general conclusions drawn of the present study are not strongly affected by the particular
parameter values.

All model parameters could, in principle, be considered to vary with time in order to account for
the observed variability. Changes in the injection rate,   
of the maximum electron energy, and  of the Doppler factor are usually assumed while modelling flares (e.g. \citealt{coppiaharonian99, sikora01,
krawczynski02, mastmoraitis08, mastetal13}), as their impact
on the SED is more direct. Other possibilities include variation of the magnetic field strength (e.g. \citealt{mastkirk97, moraitismast11}), which
might even lead to a change in the Compton dominance of the emission region in the context of SSC models.
Here we test the hypothesis that the observed marginal variability in the low state
is caused by variations of only two parameters, namely the maximum electron energy $\gamma_{e,\max}$
and the injection compactness of electrons (and protons) $\ell_{e,p}^{\rm inj}$. 
Their temporal profile must be then determined, and for this, one can adopt
one of the following approaches: (i) the {\sl phenomenological} approach, where
the functional form of the parameters is linked to the observed variability pattern
in one or more energy bands or (ii) the {\sl theoretical} approach, where
the variations are based upon a physical model for the acceleration and injection
of particles in the emission region. In this work, we adopt the first approach which
is after all the most commonly used.

 In  all three models, we assume that the variations imposed on
$\gamma_{e,\max}$ of the first component are a scaled version of the observed X-ray variability.  
This hypothesis is physically motivated by the spectral hardening observed during episodes of increasing flux (see e.g. 
Fig.~1 in A09 and Fig.~2 in \citealt{almeida14}). 
Given the X-ray flux measurements at times $t_{\rm obs, i}$, which
are transformed in the  comoving frame ($\tilde{t}_{\rm i}$) using the Doppler factor value determined at step (1),
we determine $\gamma_{\rm e, \max}$ at a previous time\footnote{The same procedure was
followed for all parameters.} $t_{\rm i}=\tilde{t}_{\rm i}-\Delta t$. Here, $\Delta t$ expresses the time in which
the photon density in the emission region reacts to
changes in the electron distribution and depends, in general, on the cooling
and escape timescales of electrons. In what follows, we set $\Delta t = t_{\rm cr}$. We verified that
slightly different values of $\Delta t$ do not alter the results. 
The intermediate values of $\gamma_{\rm e, \max}$ were then obtained by using a cubic spline interpolation scheme,: 
\eqb
\gamma_{\rm e, \max}(\tau)&=&\left \langle \gamma_{e,\max} \right \rangle
\left(\alpha_1 \frac{F_{\rm X}(\tau)}{F_{\rm X}^{\max}}\right)^{\beta_1},
\label{gemax}
\eqe
where $\left \langle \gamma_{e,\max} \right \rangle$ is the value
corresponding to the initial steady state; -- henceforth, the same will hold for all quantities enclosed by $\langle \cdots \rangle$.
In the above expression $F_{\rm X}^{\max} = 9.8\times 10^{-11}$ erg cm$^{-2}$ s$^{-1}$ is the maximum flux in the 2-10 keV energy band and $F_{\rm X}(\tau)$ is 
the result of a cubic spline interpolation of the observed X-ray lightcurve expressed in terms of the dimensionless comoving time 
$\tau=t/t_{\rm cr}$. We set the day MJD~54703.5 as the zero time ($\tau=0$) of our simulations.

In the 2-SSC model
we assume that the variability in the optical and VHE gamma-rays 
is caused by changes in the maximum electron energy of the second component.
Alternatively, one could argue that 
the optical/TeV variability can be explained by changes in the injection rate of electrons,
but this would induce non-negligible 
GeV variability, which would contradict the observations, and so we do not consider this scenario further.
Thus, $\gamma_{\rm e, \max}$
for the second component is modelled as
\eqb
\gamma_{\rm e,\max}(\tau)&=&\left \langle \gamma_{e, \max} \right \rangle
\left(\alpha_2 \frac{F_{\rm opt}(\tau)}{F_{\rm opt}^{\max}}\right)^{\beta_2},
\label{gemax2}
\eqe
where $F_{\rm opt}^{\max} = 1.64 \times 10^{-10}$ erg cm$^{-2}$ s$^{-1}$ is the maximum flux measured in the BV filters.
We use  subscripts $1,2$  for the parameters $\alpha,\beta$ to refer
to the first and second components, respectively. 
We note that if the injection compactness is kept fixed
while $\gamma_{e,\max}$ varies according to Eqs.~(\ref{gemax}) or (\ref{gemax2}), 
then the injection rate of particles is also variable (see Eqs.~(\ref{qinj}), (\ref{linj}) and (\ref{Luminj})).

Contrary to $\gamma_{e,\max}$ that varied in all three models, 
the injection compactness of primary particles was assumed to be time-dependent only in the LHs scenario. 
In this case, the emission from the optical up to X-rays 
is the result of electron synchrotron radiation, whereas the proton synchrotron component dominates in the 
gamma-ray energy band. 
Because of the tight correlation between the optical and TeV fluxes we chose to model 
$\ell_e^{\rm inj}$ and $\ell_p^{\rm inj}$ according to the  variability pattern observed in the optical filters:
\eqb
\ell_{e,p}^{\rm inj}= \left \langle \ell_{e,p}^{\rm inj} \right \rangle
\left(\frac{1}{f_{e,p}} \frac{F_{\rm opt}}{F_{\rm opt}^{\max}} +g_{e,p}\right).
\label{lelp}
\eqe
In total, we introduced eight free parameters, namely $\alpha_{1,2}$, $\beta_{1,2}$, $f_{e,p}$ and $g_{e,p}$, 
to account for the observed variability. All the parameter values, including those leading to the steady state solution 
that served as an initial condition for our calculations are summarized in Table~1.

\section{Results}
\label{results}
In the following paragraphs we present snapshots of the SED and lightcurves for each of the models discussed 
in the previous section and comment also on the pros and cons of each model.
In all cases, the very high energy (VHE) part of the derived gamma-ray
spectra has been absorbed using the 
EBL model (Model C) by \cite{finkeetal10}.
\begin{table}
\centering
\caption{Parameter values for each of the three models discussed in text.}
\label{tab-1}  
\tabcolsep=0.11cm
 \begin{threeparttable}
\begin{tabular}{ccc cc}
\hline
Parameters for& 1-SSC & LHs & \multicolumn{2}{c}{2-SSC}\\
initial steady state & \phantom{a} &  \phantom{a}& 1st & 2nd\\
\hline
\hline
$B$ (G) & 0.5 & 40 & 20 & 0.1 \\
$R$ (cm) &$10^{16}$ &  $10^{16}$ & $3\times10^{15}$ & $ 4.5\times10^{16}$\\
$\delta$ & 34 & 28 & 18 & 34 \\
\hline
$\gamma_{e,\min}$\tnote{a} & $10^{3.6}$  &  $10^{3.0}$&  $10^{3.8}$ & $10^{3.6}$  \\
$\gamma_{e,\max}$ & $10^{5.3}$ & $10^{4.7}$& $10^{4.8}$ & $10^{4.1}$ \\
$\ell_{e}^{\rm inj}$&  $10^{-4.3}$ & $10^{-4.35}$& $10^{-3.4}$ & $10^{-4.75}$  \\
$s_e$ & 2.4 &  2.6& 2.7 & 1.7 \\
$q_e$ & 1.0 & 2.0 & 1.0 & 1.0\\ 
\hline 
$\gamma_{p,\min}$ & -- & $10^{7}$& -- & --    \\ 
$\gamma_{p,\max}$\tnote{b}  & -- &  $10^{9.9}$& -- & --   \\
$\ell_p^{\rm inj}$ & -- & $10^{-6.4}$& -- & --   \\
$s_p$& -- &  2.4 & -- & -- \\
$q_p$ & -- & 2.0 & -- & --\\
\hline
Parameters for  & \multicolumn{4}{c}{\phantom{a} } \\
 variability &  \multicolumn{4}{c}{\phantom{a} }\\
\hline
\hline
$\alpha_1,\beta_1$ &  1.8,1.0 & 2.0, 1.8 & \multicolumn{2}{c}{ 2.1, 1.8}   \\
$\alpha_2,\beta_2$ & -- & -- & \multicolumn{2}{c}{2.5,1.1}\\
$f_e,f_p$ & --& 0.6, 0.6 &\multicolumn{2}{c}{--} \\
$g_e,g_p$& -- & 0.0, -0.6 & \multicolumn{2}{c}{--} \\
\hline
\hline
\end{tabular}
\begin{tablenotes}
 \item[a] These values correspond to the quantities enclosed by $\langle \cdots \rangle$ in text.
 \item[b] This value satisfies the Hillas criterion, i.e. $\gamma_{p,\max} < eBR/\mpr c^2 \simeq 1.5 \times 10^{11}$.
  \end{tablenotes}
 \end{threeparttable}
\end{table}
\subsection{One-component SSC model (1-SSC)} 
\label{ssc}
\begin{figure}
\centering
 \includegraphics[width=0.4\textwidth]{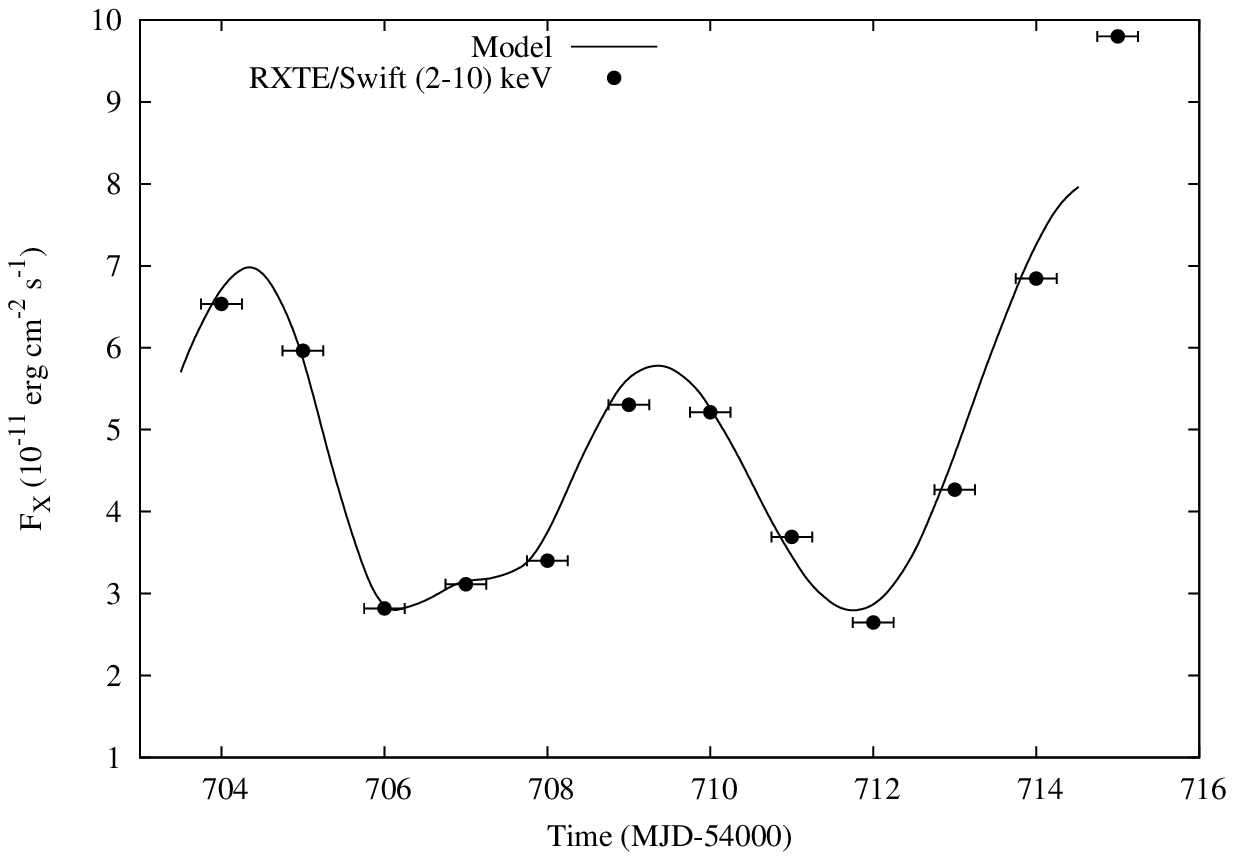}\\
 \includegraphics[width=0.4\textwidth]{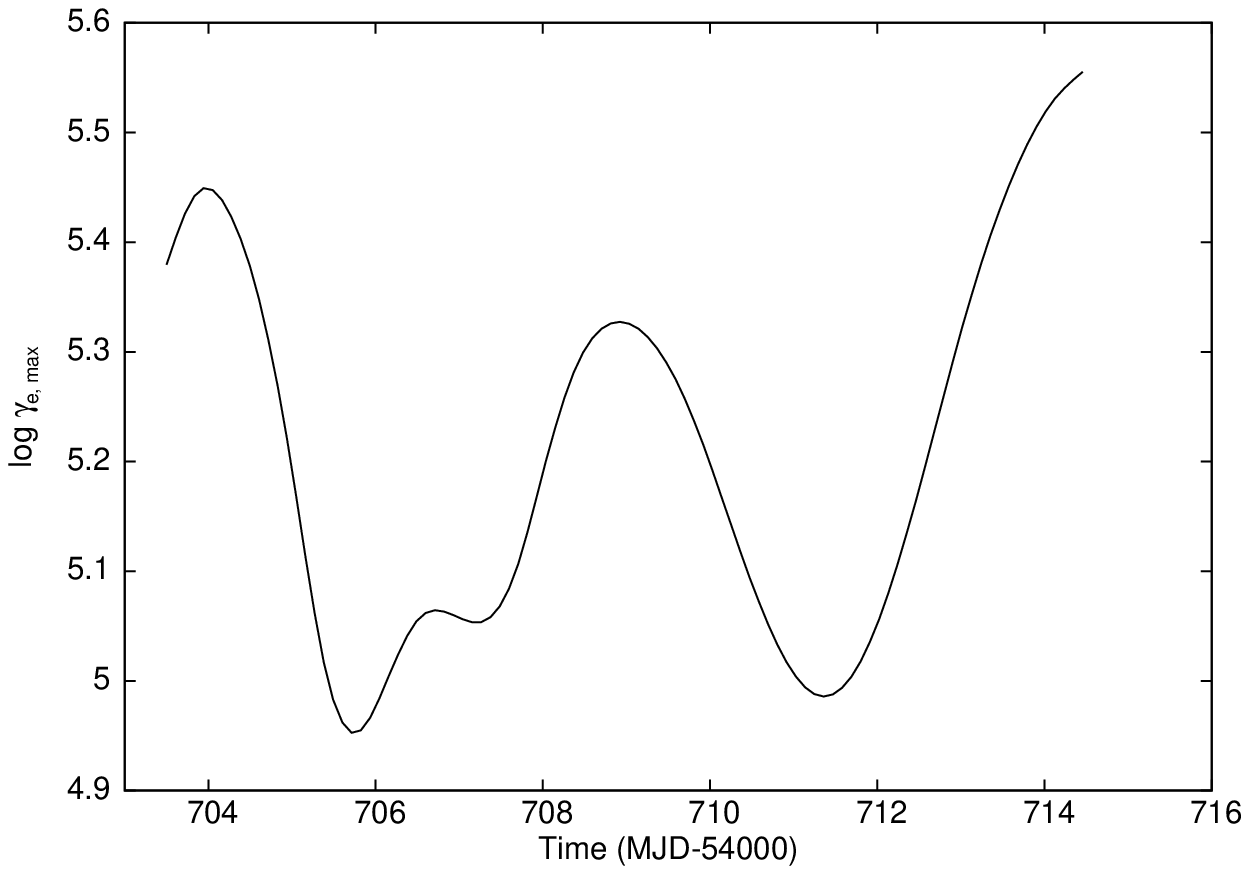}
\caption{Top panel: 
X-ray (2-10 keV) lightcurve from the 2008 campaign (A09) along with the SSC model fit. Bottom panel:
variations of $\gemx$ used as an input to the numerical code.}
\label{fig1}
\end{figure}

Figure \ref{fig1} shows the X-ray lightcurve (top panel) derived
from the model for a variable $\gamma_{e,\max}$ (Eq.~(\ref{gemax})) with
$\alpha_1=1.8$ and $\beta_1=1.0$ (bottom panel). This choice of parameters
leads to small amplitude variations of $\gamma_{e,\max}$, i.e. it varies at most by a factor of 4
 between MJD 54706 and MJD 54715 .
 
The X-ray observations (points) are satisfactorily reproduced by the model except for
the flux at MJD 54715, which is higher by a factor of $1.22$ compared to our model.
At the same date the X-ray spectrum of \pks \ was found to be harder with respect to the previous days, having a 
photon index $\Gamma_X\simeq 2.3$ (A09). Our model spectra, on the other hand, have a photon index of $\sim 2.4$.
Thus, the ratio of the observed and the model derived X-ray fluxes in the range $2-10$~keV is $\sim 0.4/0.3$, i.e. close
to the difference seen in Fig.~\ref{fig1}. If we had allowed the power-law index $s_e$
to vary with time in addition to $\gamma_{e,\max}$, the fit of the X-ray lightcurve would have been improved. 
We do not consider this case here because the number of free parameters is already large
and none of the general conclusions of the analysis would be altered.
This is a general remark that applies to the other models, too.

\begin{figure}
\centering
 \includegraphics[width=0.4\textwidth]{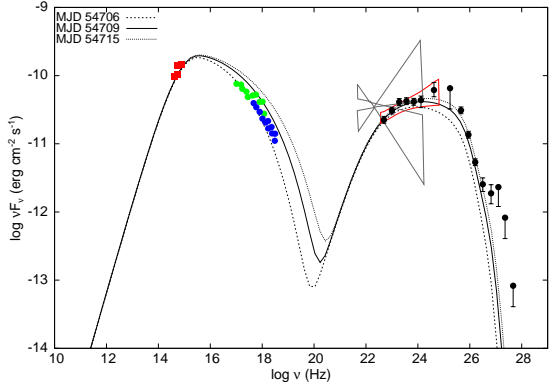} 
\caption{Snapshots of multiwavelength (MW) spectra during the period
54704-54715 MJD in the context of the one-zone SSC model. All data points are from A09. From low to high energies:
optical measurements (red points) from ATOM, combined RXTE and {\sl Swift} X-ray measurements (green \& blue points),
and gamma-ray observations (black points) by {\sl Fermi}-LAT and H.E.S.S. in the GeV and TeV energy bands, respectively. The
red butterfly is the actual {\sl Fermi} spectrum for the period MJD~54704-54715 and the grey ones show EGRET measurements.
}
\label{fig2}
\end{figure}

Snapshots of the photon spectra corresponding to MJD~54706, 54709, and 54715 
are shown in Fig.~\ref{fig2}, where all data points are taken from A09.
\begin{figure}
\centering
 \includegraphics[width=0.4\textwidth]{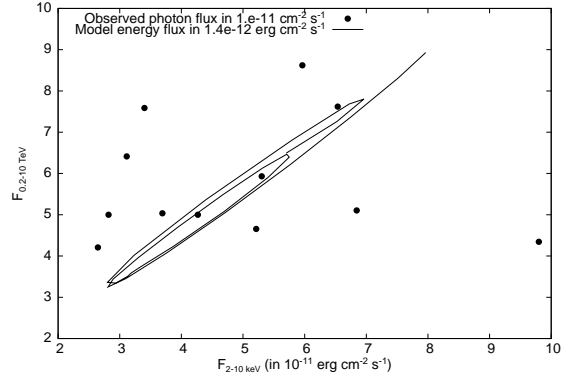} 
\caption{Plot of the gamma-ray (energy or photon) flux in 0.2-10 TeV against
the (2-10)~keV X-ray flux. 
The observations are shown with symbols, whereas the result of the 1-SSC model is plotted
with a solid line. As TeV flux, we use the observed photon flux normalized to $10^{-11}$~cm$^{-2}$~s$^{-1}$ and
the model derived energy flux in 0.2-10 TeV.
}
\label{cor}
\end{figure}
Although the model describes fairly well the X-ray behaviour,
it fails to reproduce the TeV and optical variability, 
which can be seen already from the SED snapshots in Fig.~\ref{fig2}.
Electrons emitting synchrotron radiation in the X-ray energy band lie
close to the upper cutoff of the distribution, i.e. $\gamma_X \simeq 10^5$ for $\epsilon_X=2$~keV, $\delta=34$ and
$B=0.5$~G. The same electrons will upscatter the X-ray synchrotron photons in the Klein-Nishina regime, 
since $\gamma_X \epsilon_X /( \delta m_e c^2) \gg (3/4)$. 
Thus, the upscattered photons will be produced at TeV energies, i.e. $\epsilon_{\gamma} \simeq \delta \gamma_X m_e c^2 \simeq 1.7$~TeV,
and strong correlation between the  TeV and X-ray fluxes is expected.

This is exemplified in Fig.~\ref{cor} where we plot
the TeV flux against the X-ray flux  obtained by our model (line) and by
the observations (symbols). For the plot, we used the photon flux observed
in 0.2-10 TeV with H.E.S.S. and the energy flux calculated by the model. These
are normalized to the values marked on the plot. The plot shows that the observed
X-ray and TeV fluxes are not correlated, in contrast to the model prediction. By applying a Pearson's correlation test
to the model derived fluxes we find, indeed, a strong correlation with 
the correlation coefficient being $r=0.99$. This is a robust prediction
of the 1-SSC model, which contradicts the observed loose correlation.
We find also that the TeV flux scales linearly to the X-ray one in agreement
to the analysis by \cite{katarz05}. The authors showed that the relation 
between the X-ray and TeV fluxes above the respective peaks of the SED is usually less than quadratic.

%

Moreover, the variation of $\gemx$ alone cannot account for the optical variability. 
For the adopted parameters, the cooling Lorentz factor of electrons
is $\gamma_{e, c} \approx 9\times 10^3 > \gamma_{e, \min}$ and their 
typical synchrotron frequency $\nu_{c}\simeq 4\times10^{15}$~Hz.
Thus, it is the low-energy part of the synchrotron spectrum ($F_{\nu} \propto \nu^{1/3}$) that falls in the optical window and
is unaffected by the small-amplitude variations induced at the high-energy part of the synchrotron spectrum.
One could think of a scenario where the electron injection compactness would vary according to the optical variability.
In this case, however, a larger variation of $\gemx$ would be required. This can be understood as follows:
whenever the X-ray flux increases significantly, the optical emission
is at a low level. Thus, in order 
to compensate for the low $\ell_{e}^{\rm inj}$
value, one would have to assume larger variations to $\gemx$. As a result, the 
TeV flux would show the same temporal pattern as the X-rays.
For large enough variations of $\gemx$, the model would predict 
spectral variability in the TeV energy band even in the presence of
Klein-Nishina cutoff effects, which are once again
excluded from the observations.  We do not investigate in any more
detail this hypothesis, since the X-ray/TeV correlation would still be present.

Summarizing, the one-component SSC model can explain both
the time-averaged SED and the X-ray variability despite its simplicity and the small
number of free parameters. However, its main drawback is the robust prediction of a 
tight correlation between the X-rays and TeV gamma-rays.
Interestingly, the usual SSC model was also challenged, for different reasons, by
an exceptional gamma-ray flaring event of PKS~2155-304 in 2006 \citep{aharonian09b}.

\subsection{Proton synchrotron model (LHs)}
\label{LHs}
In this scenario the low energy bump (optical up to X-rays)
is attributed to synchrotron radiation of relativistic electrons,
whereas the high-energy bump (GeV up to TeV gamma-rays) is considered
to be the synchrotron emission of a relativistic proton population. Although
the emission over the whole energy range originates from the same region, the LHs model
can be thought of as a two-component emission model because of the two independent particle populations contributing
to the low- and high-energy parts of the spectrum.
The number of free parameters required for modelling
steady-state photon emission increases from eight in the 1-SSC model to thirteen (see also Table~1).
To these one has to include six additional parameter ( $\alpha_{1,2}$, $f_{\rm p,e}$ and $g_{\rm p,e}$ )
for modelling the variability in the optical, X-rays and TeV gamma-rays.
\begin{figure*}
\centering
\begin{tabular}{c c c}
 \includegraphics[width=0.3\textwidth]{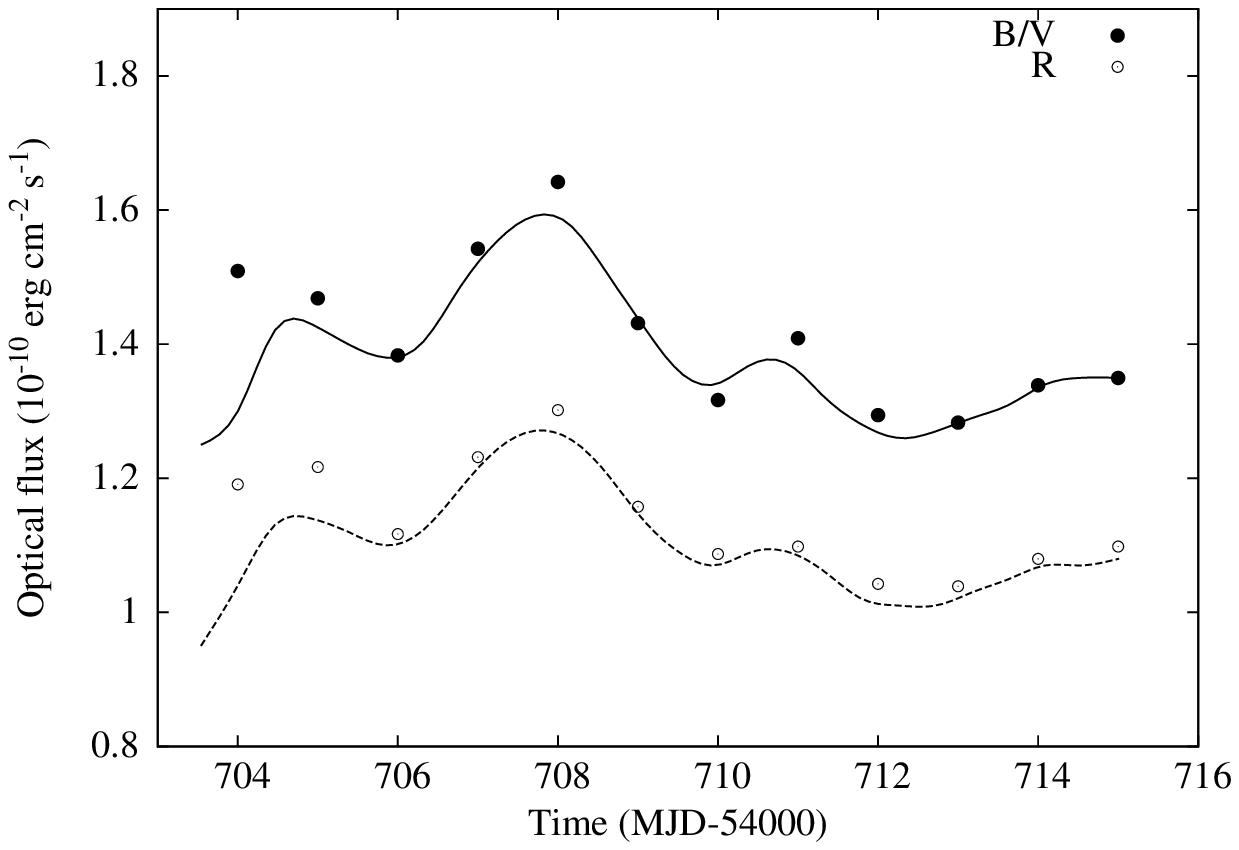} &
 \includegraphics[width=0.3\textwidth]{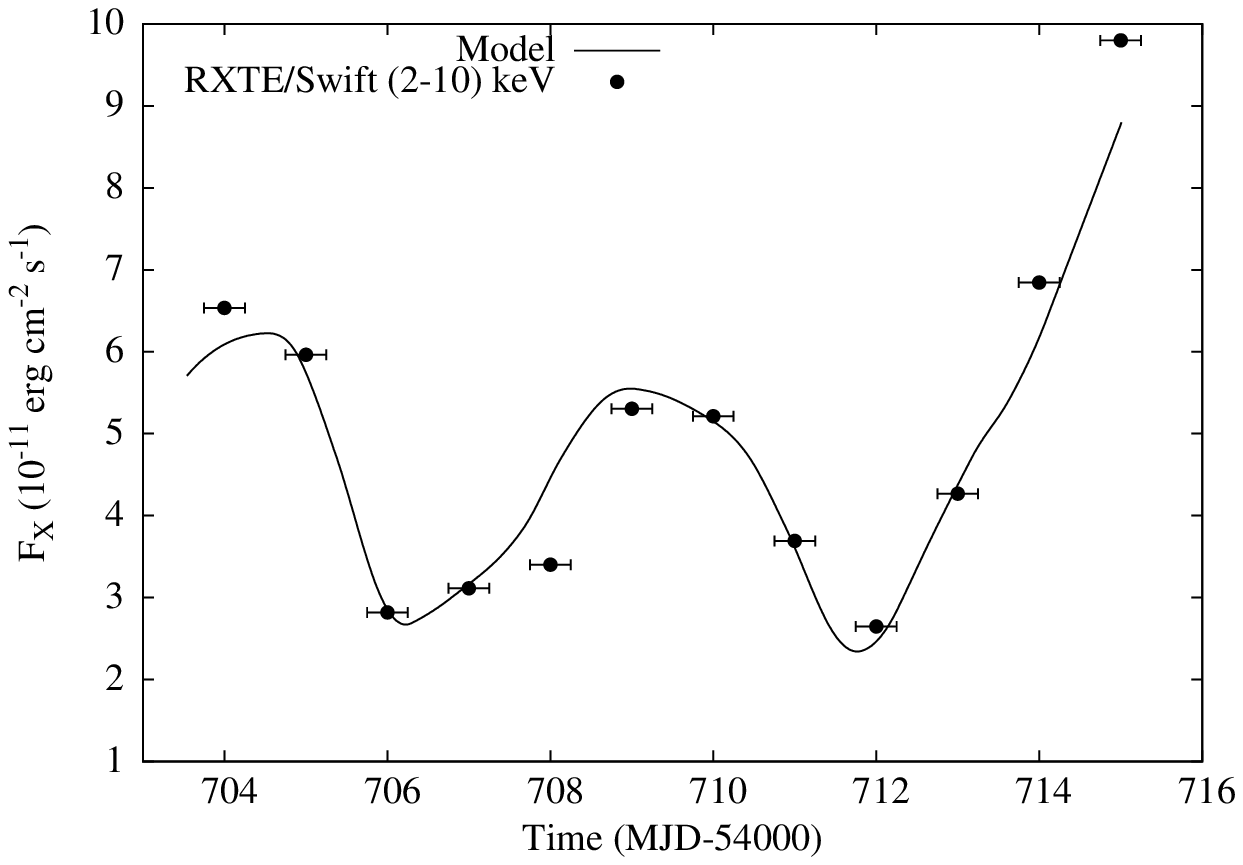} &
\includegraphics[width=0.3\textwidth]{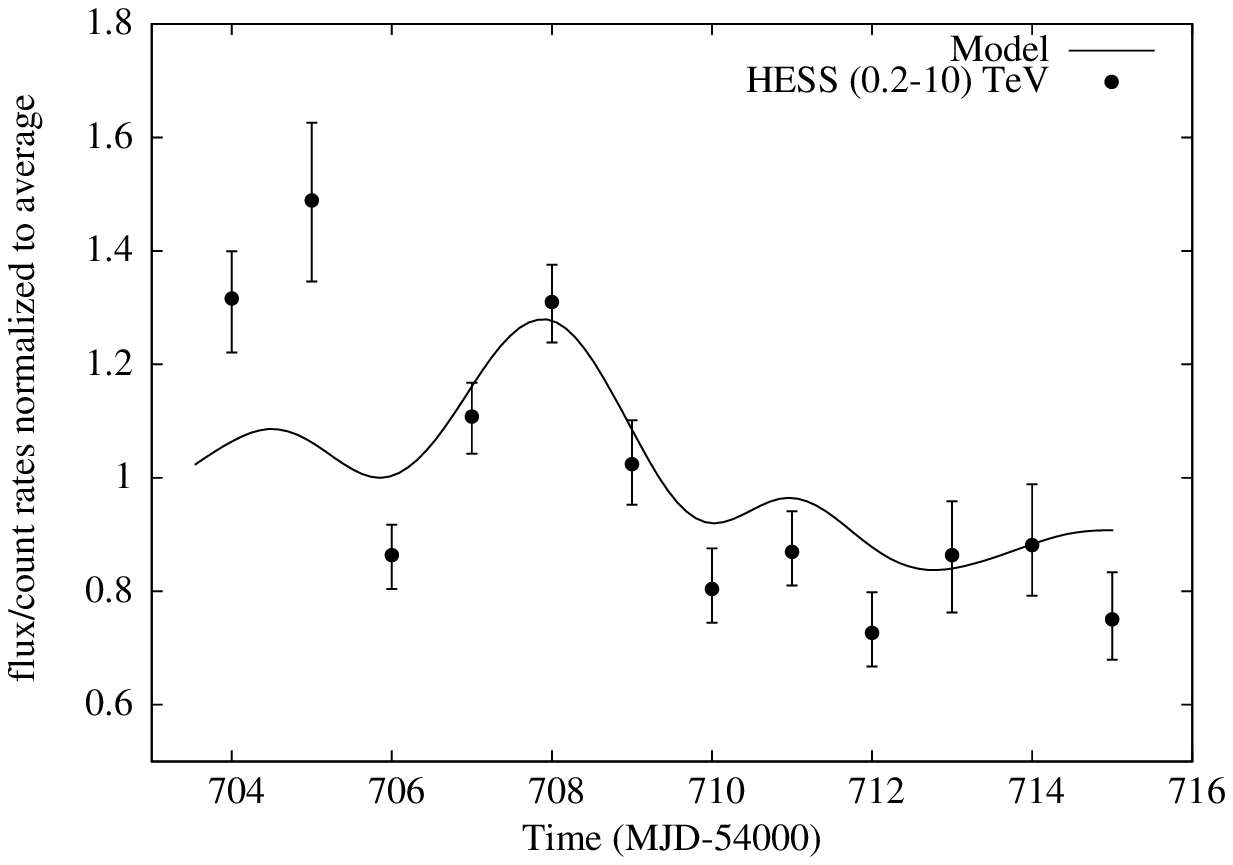} 
\end{tabular}
\caption{From left to right:  optical, X-ray and TeV lightcurves calculated
using the LHs model. For the parameters see Table~1.}
\label{lc-lhs}
\end{figure*}
In panels of Fig.~\ref{lc-lhs} from left to right we present the observed lightcurves (points)
along with the model results (solid lines) for the optical, X-ray and TeV energy
bands, respectively. 
The H.E.S.S. lightcurve in A09 was given in units of  photon count rate per effective area of the detector, and
for this reason a direct comparison to  the flux calculated with our model could not be made.
Given that we do take into account the time-averaged SED in our analysis, it is sufficient
to compare the
relative variations, i.e. $F_{\rm TeV}/\bar{F}_{\rm TeV}$ to $\dot{N}_{\rm TeV}/\bar{\dot N}_{\rm TeV}$, where 
quantities with a bar denote the average values over the twelve day period and
$\dot{N}_{\rm TeV}$
is the observed photon count rate
per effective area (see also Fig.~1 in A09). 
The model lightcurve can reproduce the observed variations apart from the first two data points of H.E.S.S.
This is not unexpected, since we modelled $\ell_{e}^{\rm inj}$ according to the optical variability, which does not correlate with the TeV lightcurve
at least for the first two days (MJD 54704-54705).

We note also that the above comparison 
is possible only in the absence of spectral variation. For this, we did not attempted a similar comparison to the Fermi
data that exhibit spectral changes, especially in the first three days, although the flux remains approximately constant. 
However, we calculated  the $0.2-300$ GeV flux of the model and found that it varies at most by a factor of 1.6, i.e. it is compatible with
a constant value. Figure~\ref{sed-lhs} shows snapshots of the SEDs obtained in the LHs scenario, where it becomes evident 
that the GeV part of the SEDs shows no spectral variation. 
To account for the observed spectral variations in the GeV band,
one should treat an additional model parameter, e.g. the power-law index $s_p$ of the
proton distribution, as time dependent.

\begin{figure}
\centering
\begin{tabular}{c}
 \includegraphics[width=0.35\textwidth]{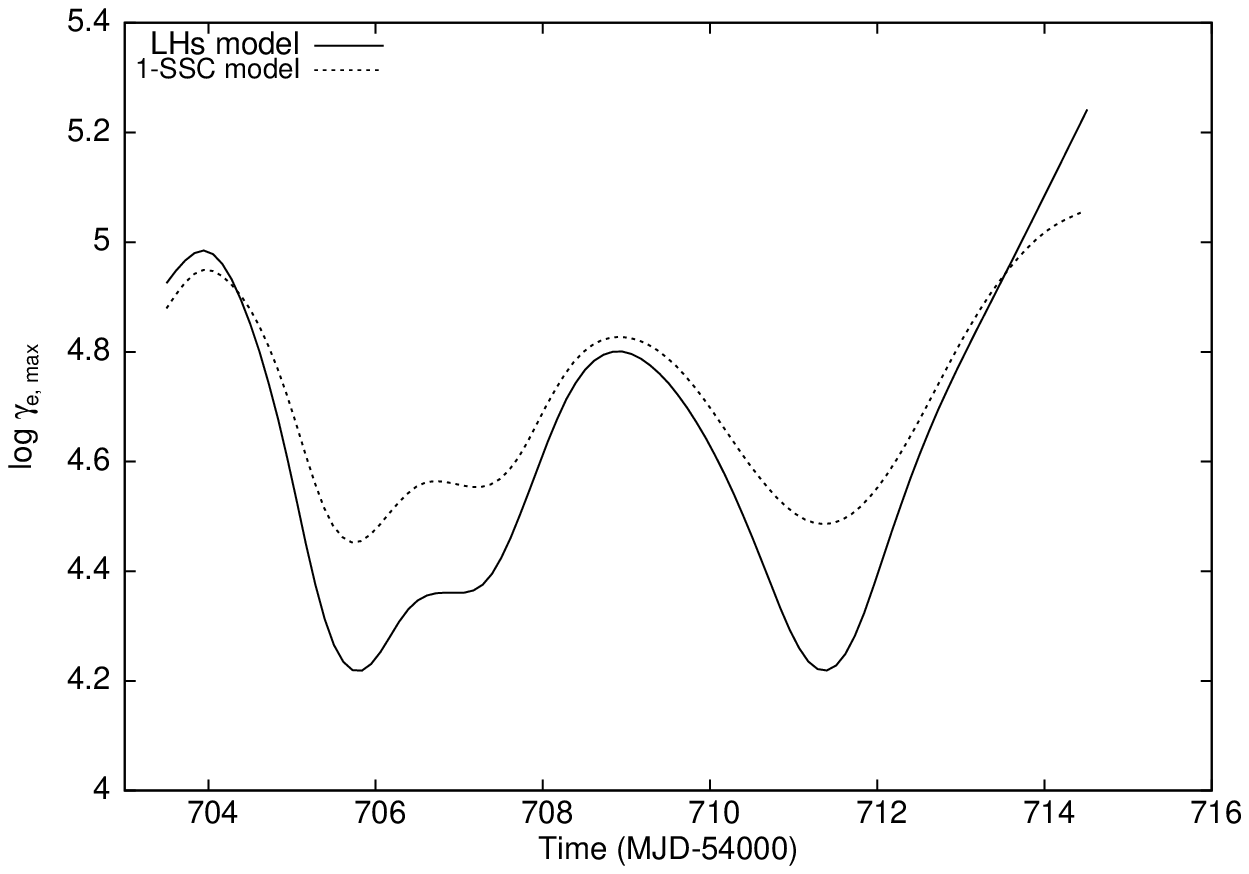} \\
 \includegraphics[width=0.37\textwidth]{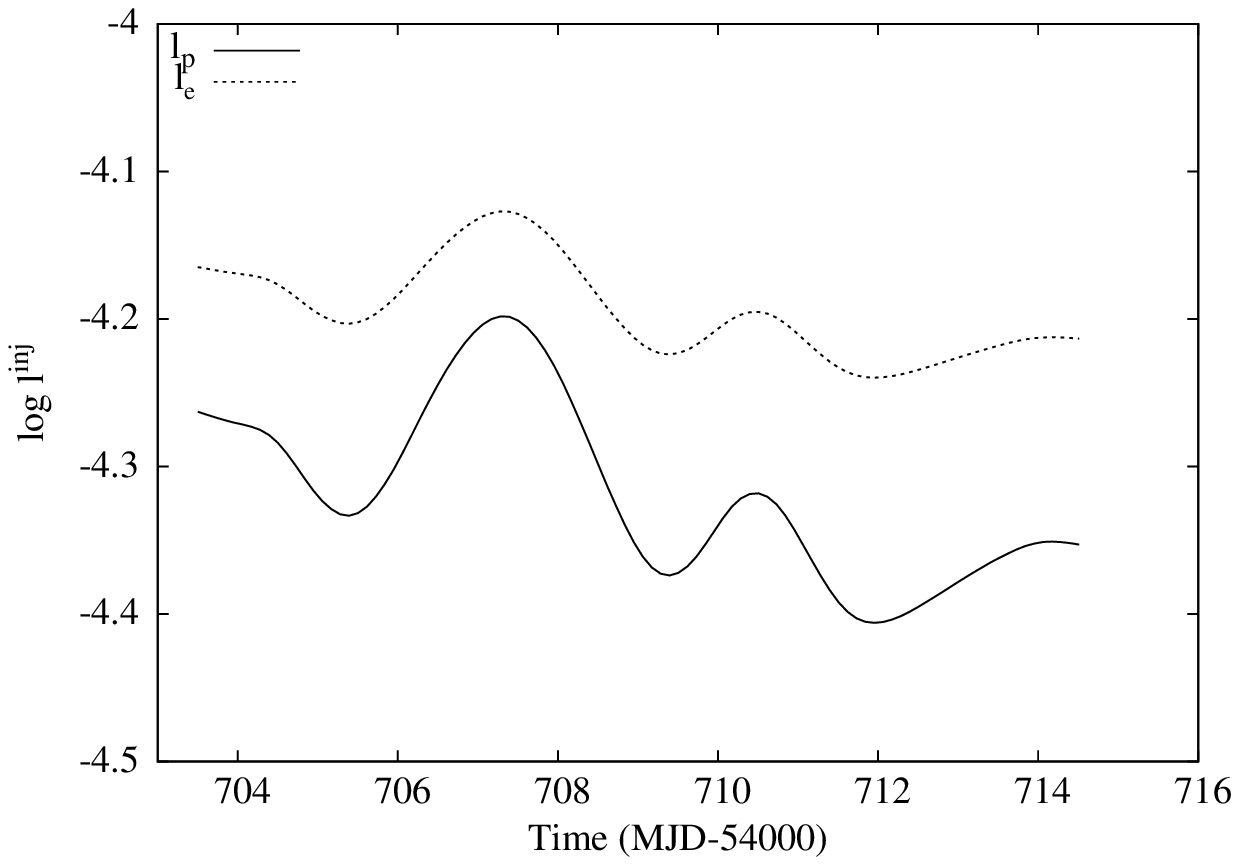}  
\end{tabular}
\caption{Top panel: comparison of $\gamma_{e, \max}(\tau)$ used for the LHs (solid line) and 1-SSC (dashed line) models;
for comparison reasons, the latter is shifted by a factor of -0.5 in logarithm. Bottom panel: plot of $\ell_p^{\rm inj}$ (solid line) and $\ell_e^{\rm inj}$ (dashed
line) as a function of time where the former is shifted by +2.3 logarithmic units.}
\label{input-lhs}
\end{figure}

\begin{figure}
\centering
 \includegraphics[width=0.4\textwidth]{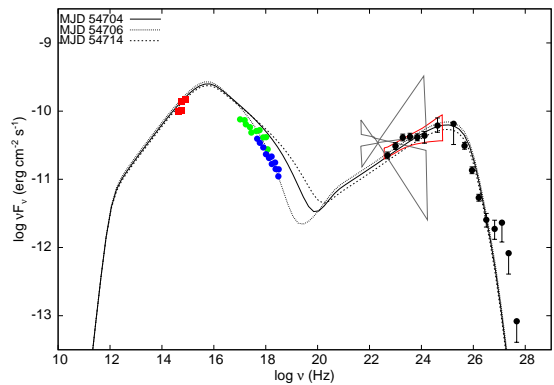} 
\caption{Snapshots of the SED obtained in the context of the LHs model 
for MJD 54704, 54706 and 54714.}
\label{sed-lhs}
\end{figure}

The above results were obtained by varying $\gamma_{e,\max}$ and $\ell_{e,p}^{\rm inj}$ 
as shown in the top and bottom panels of Fig.~\ref{input-lhs}.
The relative change of $\gamma_{e,\max}$ in the LHs model (solid line)
is larger compared to that required by the  1-SSC model (dashed line), mainly because
of the steeper electron distribution assumed ($s_e=2.6$). 
Besides $\gamma_{e,\max}$, which changes by an order of magnitude,  the variations
of the injection compactnesses are small. 

Summarizing, the LHs model can account for (i) the observed flux variability
from the optical up to the TeV energy band; (ii) the approximately constant GeV flux; and 
(iii) the absence of correlation between the X-rays and TeV gamma-rays but it cannot explain the spectral variability in GeV
gamma-rays unless the proton distribution slope
changes. In total, it requires three (or four) model parameters to be functions of time.

\begin{figure}
\centering
  \includegraphics[width=0.4\textwidth]{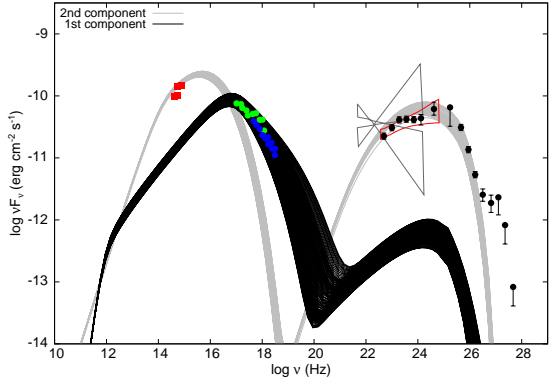}
\caption{MW photon spectra calculated using the 2-SSC model for the whole period
MJD 54704-54715 that demonstrate the difference in the range of variance of the two components.
The emission from the first and second components
 are plotted with black and grey lines, respectively. For the parameters used, see Table~\ref{tab-1}.}
\label{sed-ssc2}
\end{figure}

\subsection{Two-component SSC model (2-SSC)}
\label{2ssc}

The uncorrelated  variability between the optical and X-ray fluxes is one of the most intriguing results
of the 2008 campaign. Motivated by the difficulty to explain the presence of 
no correlation in the context of a homogeneous one-component emission model (some of the problems
were already discussed in  \S\ref{ssc} and \S\ref{LHs}), we considered a two-component SSC scenario, where
the total emission originates from two physically distinct regions.
Synchrotron radiation from the first one accounts only for the X-ray flux, whereas the emission from the second
component dominates in the optical (synchrotron) and gamma-ray  (SSC) energy bands.
The above physical setup implies that the parameters describing the first emission region should be such as 
to suppress its SSC emission. Moreover, we treat both regions independently by assuming
that only the synchrotron photons produced
within the same region serve as targets for inverse Compton scattering. For the parameters used
here, this assumption is valid as long as the separation $r_{12}$ between the two regions is a few times the size
of the larger component, i.e. $r_{12} \gtrsim 6\times 10^{16}$~cm (for more details, see Appendix).

Figure~\ref{sed-ssc2} shows 
snapshots of SEDs for both components 
covering the whole observing period. The SED of the first component (black lines) is clearly synchrotron dominated with a peak
at $\sim 0.4$ keV and a SSC peak luminosity being approximately $1\%$ of the synchrotron peak luminosity. 
The SED of the second component (grey lines) has the usual double-humped shape with the synchrotron  
and SSC spectra peaking at far-UV and $\sim$~GeV energies, respectively. Because of the underlying
electron distribution, which is hard and spans over only one decade in energy (see Table~\ref{tab-1}),
the two bumps of the SED appear narrow with large curvature (for relevant detailed discussion, see \citealt{massaro05}). 
This results in the following: the second component contributes only a small fraction to the 
2-10~keV flux, since  its  integrated X-ray flux
does not exceed the value $3\times 10^{-12}$~erg~cm$^{-2}$~s$^{-1}$ and the model slightly underpredicts 
the H.E.S.S. flux. This underprediction could be resolved by 
assuming a broader electron distribution. This would be, however, problematic, 
as the  synchrotron spectra would dominate the emission in the X-rays and thus destroy the loose
correlation between them and the VHE gamma-rays. 

 The differences in the spectral shape reflect the differences in the underlying physical quantities
describing each component (see Table~1), in contrast to \cite{almeida14} where similar values of $\delta$, $B$ and $R$ were attributed to both components.
In our analysis, the first component is more compact because of its smaller size and its lower Doppler factor and contains a stronger magnetic field.
One can relate the location $r_i$ of  each component to its radius $R_i$ as $R_i\approx r_i \theta_i$, where $\theta_i$
is the opening angle of each region, which typically is smaller than the opening angle of the jet.
In the limit where $\delta_i \simeq \Gamma_i$ and under the assumption of collimation\footnote{For a detailed discussion
with specific examples from FSRQs, see \cite{nalewajko14}.}, i.e. $\Gamma_i \theta_i < 1$,
one finds that $r_i \gtrsim R_i \delta_i$. Using the values of Table~1 we find that $r_1 \gtrsim 5.4\times10^{16}$~cm and $r_2 \gtrsim 1.5\times10^{18}$~cm,
i.e. the first and second components should be placed at the sub-pc and pc-scale jet, respectively.
We also note that the synchrotron self-absorption frequency of the first component appears at $\sim 10^{12}$~Hz, i.e. well above
the GHz radio band. Thus, one could go one step further and argue that any radio variability observed in the low state
should be correlated with the optical and VHE gamma-rays but not with the X-rays; such observational evidence, however,  is still lacking.
For a better comparison of the energetics, we summarize in Table~\ref{table2} the average\footnote{We calculate the average
value over the period MJD~54704-54715.} photon and electron energy densities together with the magnetic energy density
for the two-components as measured in their respective rest frames.
For the first and second components we find $\bar{u}_e \ll \bar{u}_{\gamma} \ll u_B$ and 
$\bar{u}_{\gamma} \lesssim u_B < \bar{u}_e$, respectively. 

Besides the differences in   their physical properties and emitted SEDs, the two components show also differences in their variability properties.
This can be seen by the clustering of all the snapshots of the second component in contrast
to the wider range of variance of the first component. To derive the above SEDs we modelled
$\gamma_{e,\max}$ according to Eqs.~(\ref{gemax}) 
and (\ref{gemax2}) for the first and second component, respectively.
In this way we can ensure once again the correlation between the optical and TeV fluxes, while
the X-ray emission will be uncorrelated with both of them. 

\begin{figure*}
\centering
\begin{tabular}{c c c}
 \includegraphics[width=0.3\textwidth]{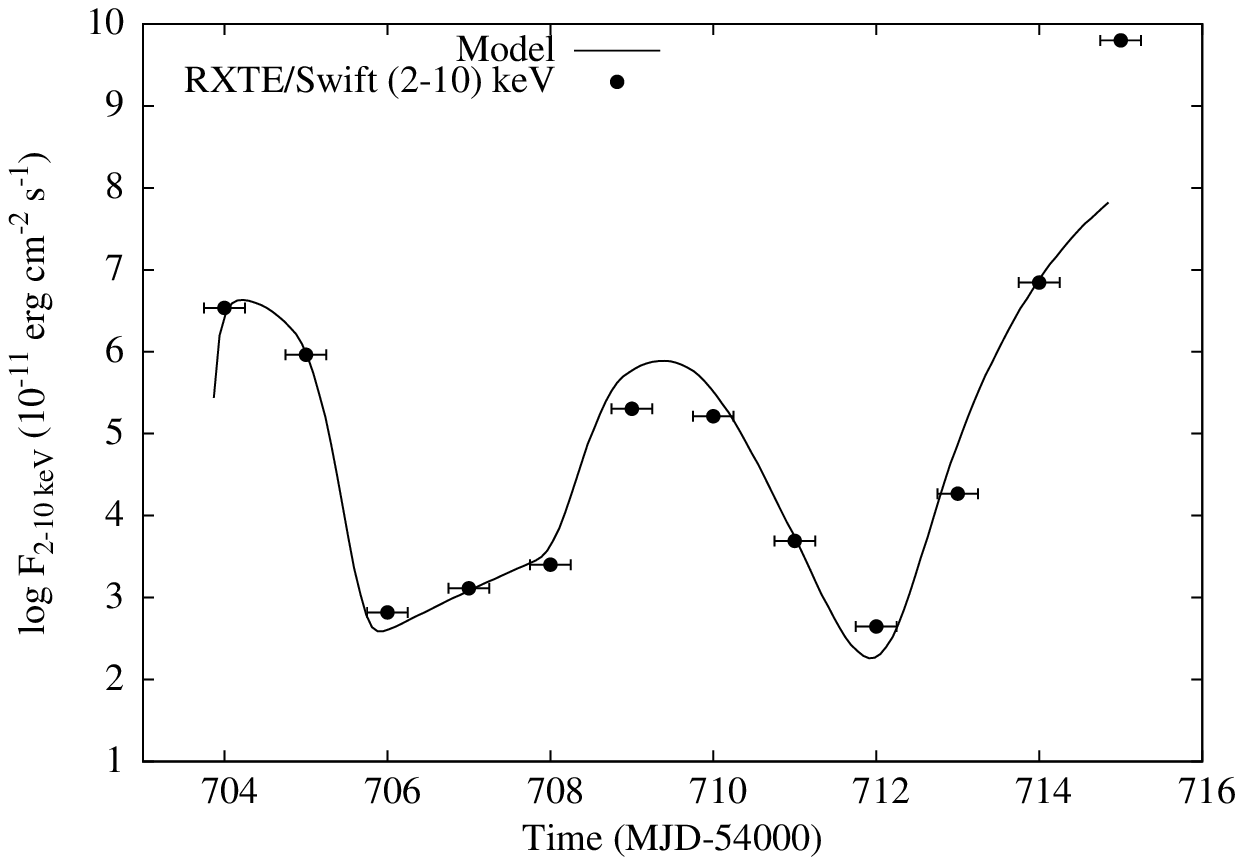} &
 \includegraphics[width=0.3\textwidth]{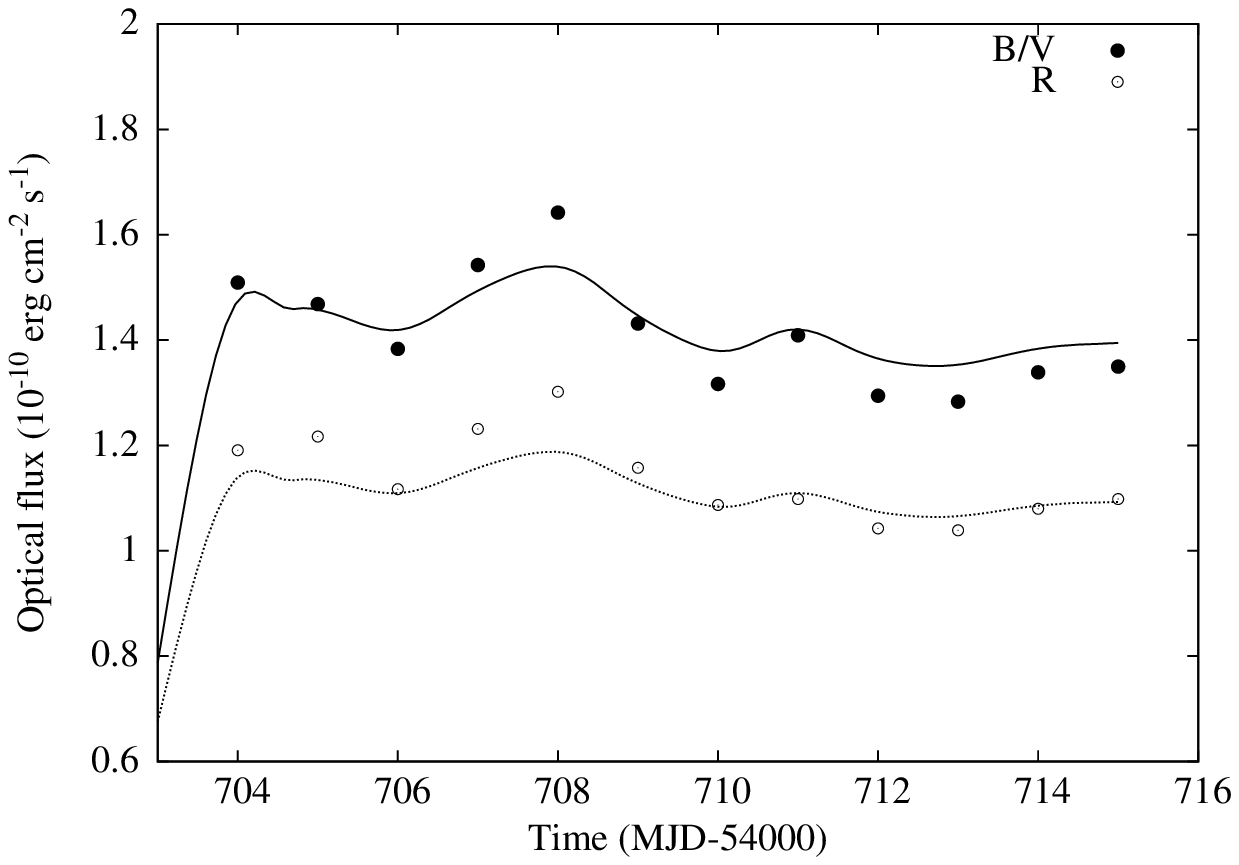} &
 \includegraphics[width=0.3\textwidth]{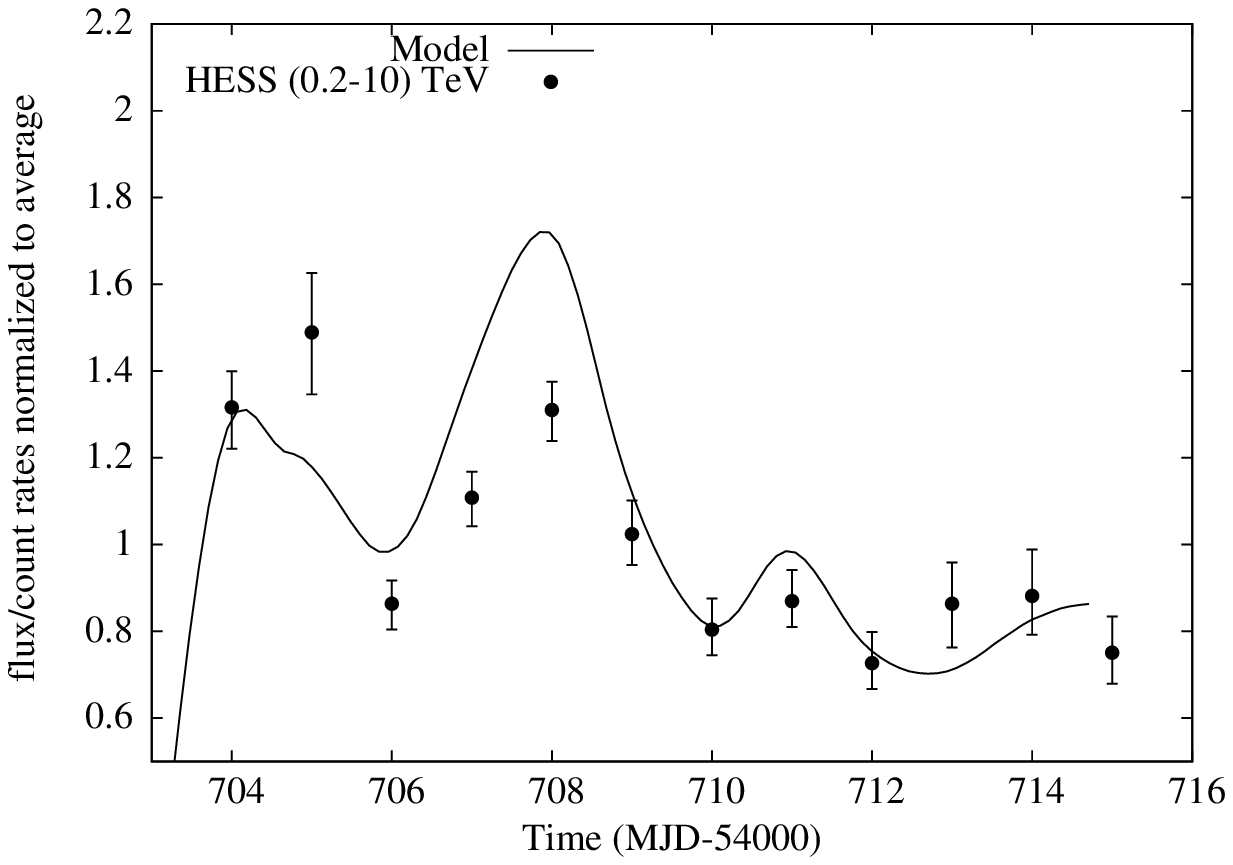} 
\end{tabular}
\caption{Same as in Fig.~\ref{lc-lhs} but for the two-component SSC model.}
\label{lc-ssc2}
\end{figure*}

The lightcurves obtained from the 2-SSC model are presented in Fig.~\ref{lc-ssc2}. The abrupt increase in the 
model derived optical and TeV fluxes just before MJD~54704 has to do with the initial steady state, and 
in this sense it could be avoided for another choice of initial conditions. For this, we do not 
consider the early part in the discussion of our results.
The TeV lightcurve is the less satisfactory result of the 2-SSC model, since it cannot
reproduce the increasing trend of the first two days, for the same reasons as in the LHs scenario,
and it overestimates the ratio $F_{\rm TeV}/\bar{F}_{\rm TeV}$ between the days 54706-54708.
We note that given the freedom that the multiparameter models provide, a more detailed search of the parameter
space might result in a better fit.

Summarizing, we showed that the 2-SSC model can satisfactorily explain (i)
the time-averaged SED; (ii) the optical and X-ray lightcurves; (iii) the optical/TeV correlation; and
(iv) the absence of correlation between the X-rays and TeV gamma-rays, yet  the model requires a large
number of free parameters and suggests 
the presence of two distinct regions, being energetically different, within the same jet, something that is difficult to reconcile with 
the present understanding of jet physics.

\begin{table}
 \centering
 \caption{Photon ($\bar{u}_{\gamma}$), electron ($\bar{u}_e$) and magnetic ($u_B$) energy densities for the 2-SSC scenario. For the first two quantities we present
 their average values over the period MJD~54704-54715.}
 \begin{tabular}{cccc}
  \hline
 \phantom{} & $\bar{u}_{\gamma}$ (erg/cm$^3$)& $\bar{u}_e$  (erg/cm$^3$) & $u_B$  (erg/cm$^3$) \\ 
 \hline
 {\rm 1st} & $4\times10^{-1}$ & $4.7\times10^{-3}$ & $16$ \\
 {\rm 2nd} & $5\times10^{-4}$ & $3\times10^{-3}$ & $4\times 10^{-4}$ \\
 \hline
 \end{tabular}
\label{table2}
\end{table}

\subsubsection{A flaring event}
\label{flares}
Here we present two indicative examples of
flaring events because of the activation of one of the emitting components in the 
context of the 2-SSC scenario. In particular, we assume that the period of low activity (MJD 54704-54715) is followed by
an active period of the second component lasting nine days.
To model the high state period we introduced Lorentzian variations to
either $\gamma_{e,\max}$ or $\ell_e^{\rm inj}$. All the parameters, except for those that have to do with
the variability, are the same as in Table~1. 

\begin{figure*}
\centering
\begin{tabular}{c c}
 \includegraphics[width=0.4\textwidth]{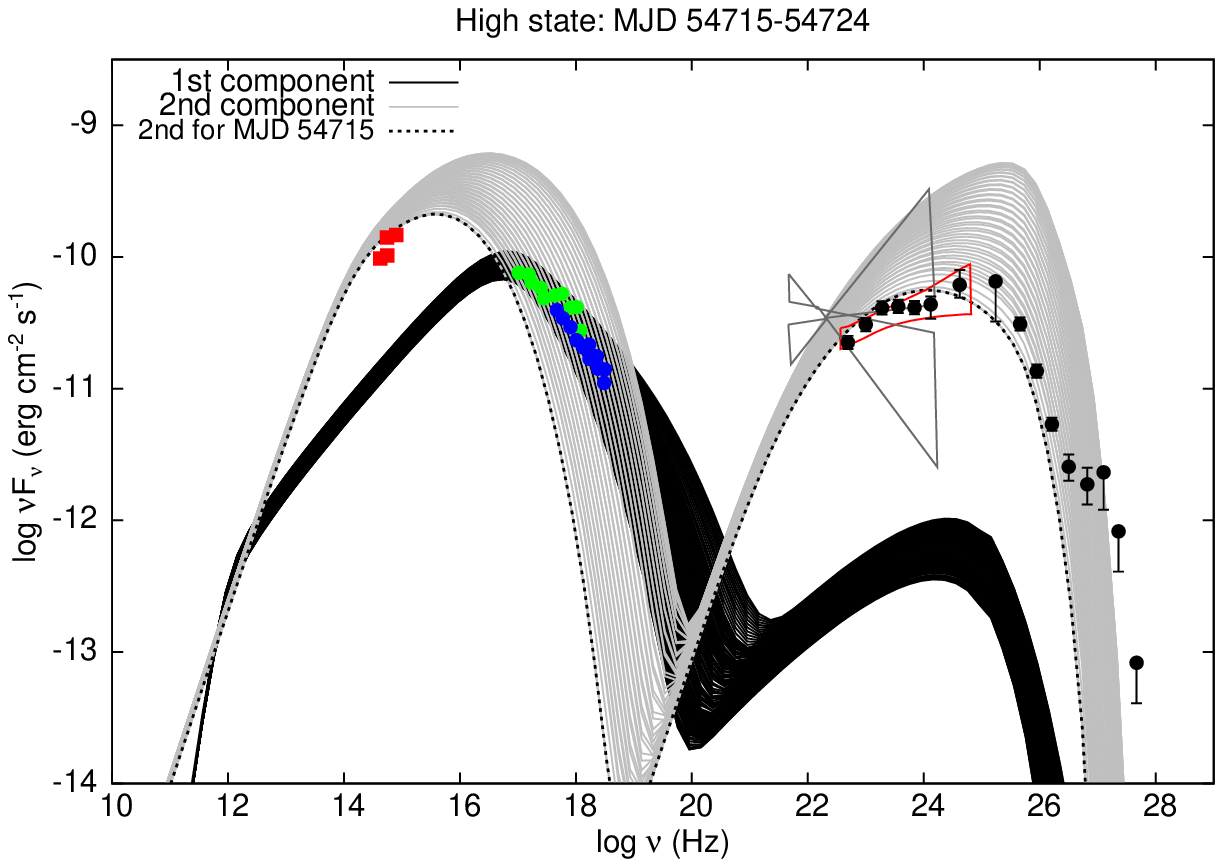} &
 \includegraphics[width=0.4\textwidth]{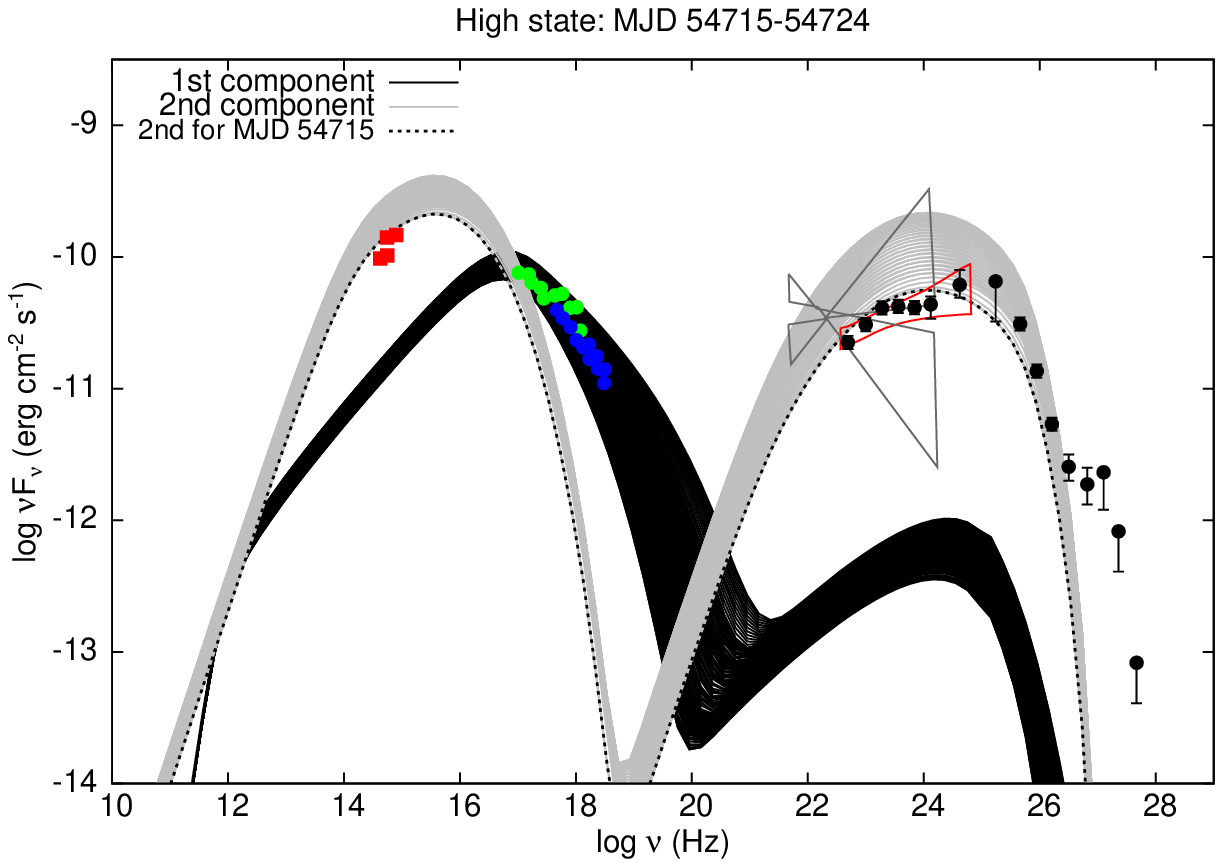} 
\end{tabular}
\caption{Time-dependent SEDs (grey lines) of a fiducial high-state period starting after MJD 54715 for the two
scenarios discussed in text. The results for Flares A and B are shown in the left and right panels, respectively.
In both panels the last SED of the low-state for the second component is also shown with black dashed line.
The first component retains the same small-amplitude variability of the low-state period (black lines). For comparison reasons,
the observations
of the low state period are also shown.
}
\label{sed-flare}
\end{figure*}

\begin{figure*}
\centering
\begin{tabular}{c c}
 \includegraphics[width=0.4\textwidth]{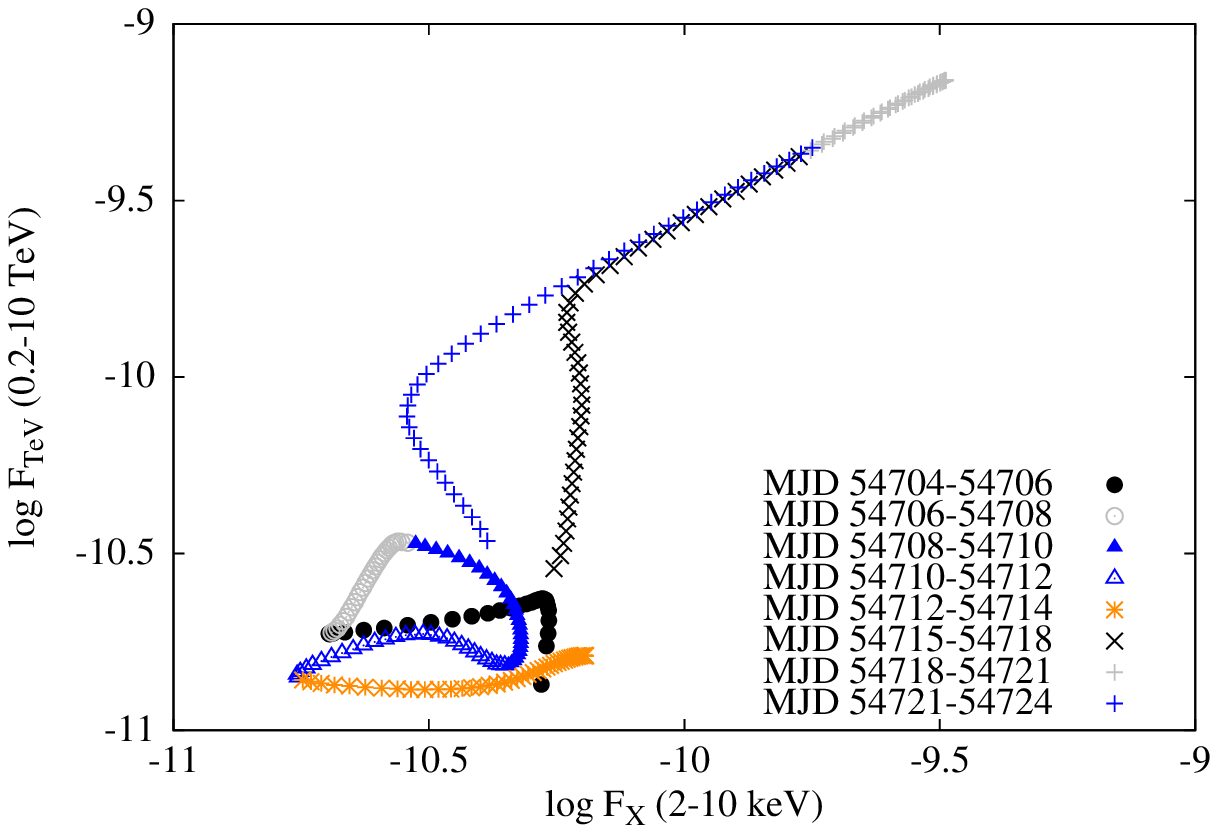} &
 \includegraphics[width=0.4\textwidth]{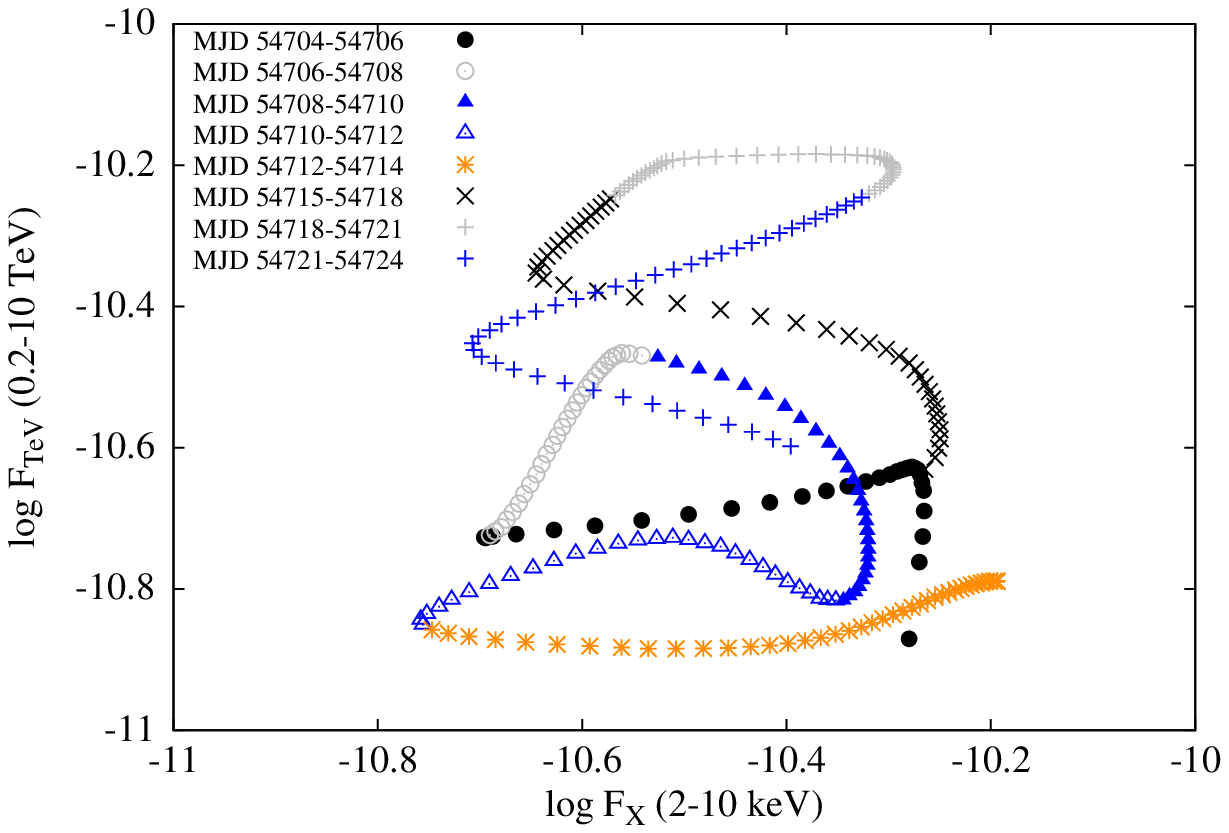} 
\end{tabular}
\caption{The $F_{\rm TeV}-F_{\rm X}$ plane in logarithmic scale for the periods of low and high activity.
The results for flares A and B are shown in the left and right panels, respectively. Different symbols 
are used to denote different phases of the lightcurve.}
\label{cor-flare}
\end{figure*}

We investigated the following cases:
\begin{itemize}
 \item Flare A: $\gamma_{e,\max}$ is varying according to 
 \eqb
 \frac{\gamma_{e, \max}(\tau)}{\gamma_{e, \max}(\tau_e)} = \frac{\tau_p^2+(G/2)^2}{(\tau-\tau_p)^2+(G/2)^2}, \ \tau \ge \tau_e,
 \eqe
 where $\tau_e$ is end time of the low state period (in $\tcr$ units), which we set equal to zero,
 $G=25$, $\tau_p=25$  and  $\gamma_{e,\max}(\tau_e) = 10^{4.5}$. 
 The increase in $\gamma_{e,\max}$ leads to correlated optical, X-ray and gamma-ray flares.
  \item Flare B:   $\ell_e^{\rm inj}$ is varying according to 
  \eqb
 \frac{\ell_e^{\rm inj}(\tau)}{ \ell_e^{\rm inj}(\tau_e)} = \frac{\tau_p^2+(G/2)^2}{(\tau-\tau_p)^2+(G/2)^2},\ \tau \ge \tau_e,
 \eqe
 where $G=50$, $\tau_p=25$ and $\ell_e^{\rm inj}(\tau_e)=10^{-4.75}$. The increase in the electron compactness results
 in a correlated optical/gamma-ray flare.
\end{itemize}
Our results are summarized in Fig.~\ref{sed-flare} where
the left and right panels show time-dependent SEDs obtained for Flares A and B, respectively.
Even though $\gamma_{e,\max}$ changes only by a factor of 4 in Flare A,
the luminosity of both the synchrotron and SSC components increases significantly, 
because of the hard energy spectrum of the electron distribution.
At the beginning of the active period, the X-ray emission is still dominated by the first component but
as $\gamma_{e,\max}$ gradually increases, both components, each of them having its own temporal behaviour,
contribute to the X-ray flux. This may lead to interesting patterns in the $F_{\rm TeV}-F_{\rm X}$ plane, as it
can be seen in the left panel of Fig.~\ref{cor-flare}. 
One can distinguish between phases of strong correlation, e.g. MJD 54718-54721, but also
periods where the two are anti-correlated. The reason is that the X-ray flux
is the sum of both components, each of them contributing the most to the total flux at different periods,
while having different lightcurve shapes.

In flare B the contribution of each component to the overall SED during the high-state period is clear.
Such scenarios could explain correlated optical and gamma-ray activity, with the amplitude of optical flares being typically
smaller than the one in gamma-rays (see e.g. right panel in Fig.~\ref{sed-flare}).
The flux-flux diagram for the low-state and high-state periods is shown in the right panel of Fig.~\ref{cor-flare}.
We find no significant correlation when we take into account the whole period of 20 days; the Pearson's
correlation coefficient is $r=-0.07$ with probability $P<24\%$ of a chance correlation.
The correlation may be enhanced, however, if we focus on subsets of 1-3 days.
Thus, even in scenarios where the TeV and X-ray fluxes 
originate from different components, each of them having its own temporal behaviour,
a strong correlation between the two may be obtained.

One could think of scenarios where the first component is the active one, thus leading
to correlated X-ray/gamma-ray variability and an approximately constant optical flux.
 For the adopted parameter values, the SSC emission of the  first component  is strongly suppressed
and changes of just $\ell_{e}^{\rm inj}$ and/or $\gamma_{e,\max}$ would not suffice to enhance
the SSC emission to the observed flux levels. Thus, this flaring scenario
would require changes in the B-field and/or size of the first emitting region and should be 
studied in more detail in a future work.

\section{Discussion}
\label{discussion}
The high quality of observational data has allowed us to attempt both spectral and
temporal fits to the  observations of blazar PKS~2155-304 while being in a low state (25 August-6 September 2008).
We applied both leptonic and leptohadronic scenarios and focused on variants that 
were successful in fitting the time-averaged SED.

 We showed that the one-component SSC model 
can explain the time-averaged SED
as well as the X-ray variability. However, it predicts a tight correlation
between the X-rays and TeV gamma-rays,  which contradicts the loose correlation
observed in the low state of \pks.~ For this reason, we suggest 
the 1-SSC model to be the least plausible among the three scenarios, although
its simplicity makes it the most attractive.
We have shown that two-component emission models, namely the LHs and 2-SSC models, are more adequite in reproducing 
the observed SED and lightcurves,  at the cost of a large number of free parameters. We note also 
that the aforementioned models have also their
share of weak points, the most important being
the disagreement between the model and observed TeV lightcurves at early times (MJD~54704-54705).
Other variants of the leptohadronic model, such as the photopion model where the gamma-ray emission is attributed
to the synchrotron radiation of pairs produced through pion decay, were not
discussed here because of their intrinsic TeV/X-ray flux correlation \citep{mastetal13}. 

 In all three scenarios, the observed marginal variability 
was modelled by assuming that the maximum energy of electrons
and/or the injection compactness of particles are variable, with the required variations being
 a scaled version of the lightcurve in one or more energy bands (see \S3).
In principle, the changes in both $\ell^{\rm inj}_{e,p}$ and $\gamma_{e,\max}$
could be related to variations of the conditions in the acceleration zone, which 
in our framework, is considered to be a black box acting
as a reservoir of accelerated particles for the emission (radiation) zone.
In the case of shock acceleration for example, the encounter of the shock
with a region of higher (lower) density could result in an increase (decrease)
in the injection rate of particles in the acceleration region, and subsequently,
in the emission region. Variations of the maximum energy
of accelerated particles usually imply changes in the acceleration ($t_{\rm acc}$) or/and energy loss ($t_{\rm loss}$) timescales, since
$\gamma_{\max}$ is the energy where the energy loss and energy gain rates become equal (see e.g. \citealt{dermerhumi01, petroetal14}).
For example, in the simplest scenario where particles are shock accelerated at the Bohm rate and lose energy through synchrotron radiation, 
particle acceleration saturates
at $\gamma_{e, \max} = \left(6\pi e / \sth B_0 \right)^{1/2}$, with $B_0$ being the magnetic field strength in the acceleration zone and $e$
the electron charge. Doubling of $\gamma_{e, \max}$ would thus require a decrease of $B_0$ by a factor of 4.
The dependence of $\gamma_{e, \max}$ on the various physical quantities, however,
is critically determined by the acceleration mechanism at work (for possible acceleration mechanisms and 
their respective $t_{\rm acc}$, see \cite{tammiduffy08} and references therein).  Thus, an interpretation
of the derived parameter variations in the context of a particular physical mechanism lies out of the scope of the present work.

The emission of blazars in  high (flaring) states is what usually draws the attention, since
modelling of high states  may give insight to the properties of the radiating particles (for a relevant discussion, see e.g. \citealt{aharonian09b}).
Our time-dependent analysis showed, however, that we can deduce information about the properties
of the emission region(s) even when the observations correspond to periods of low activity.
The particular set of observations has been also recently discussed by \cite{almeida14} in a different
context. This is a great opportunity for a qualitative comparison of the results obtained by two
different approaches for the same dataset and for the same source. Starting from a different basis, namely
optical polarization measurements during the 54711-54715 period,  \cite{almeida14} conclude 
that the X-ray emission must originate from the same component that is responsible for the variations
seen in the polarization of optical data. At the same time, this component should be hidden by the second one
both in the optical and gamma-ray energy bands -- for comparison see Fig.~\ref{sed-ssc2} in this work and
Fig.~3 in \cite{almeida14}. Given that we did not aim to derive the best-fit  parameter values, a quantitative comparison
lies out of the scope of the present study.

Although the results of our time-resolved analysis do not 
favour the LHs over the 2-SSC model, they suggest that the SED of PKS~2155-304 in a low state is composed
by the emission of at least two components, which correspond to either 
different particle populations or spatially different emission regions.
In this scenario a loose correlation between the various energy bands is expected whenever
the source is not active, whereas if the source enters a high state where only one of the components lights up,
tight correlations should be expected. Such correlations are quite common in flaring blazars, e.g. Mrk~421 \citep{fossati08}, and
this was indeed the case during the exceptional flaring event of PKS~2155-304 in 2006 \citep{aharonian09b}, where
a strong correlation between the X-rays and TeV gamma-ray was observed (for more details, see \citealt{aharonian09b}).

Concluding, the time-resolved analysis of the MW observations of 
\pks \  
between August and September 2008,  
points that the quiescent emission of a blazar may result
from a superposition of the radiation from different
components, whereas the high state emission  may  still be  the result of a single
component. This demonstrates the importance of contemporeneous monitoring of blazars and shows
 how observations in low states can be used to gain insight on the properties of the emission region.



\section{Acknowledgments}
I would like to thank Prof.~A. Mastichiadis for helpful
discussions  and the anonymous referee for his/her comments that helped to improve the manuscript.
Support for this work was provided by NASA 
through Einstein Postdoctoral 
Fellowship grant number PF3~140113 awarded by the Chandra X-ray 
Center, which is operated by the Smithsonian Astrophysical Observatory
for NASA under contract NAS8-03060.

\appendix
\section{Separation distance in the 2-SSC model}
In the 2-SSC model we treated the two components independently, i.e.
we used as seed photons for the inverse Compoton scattering
only the synchrotron photons produced within each component.
We calculate the minimum separation distance between the two components
needed for this assumption to hold.

In what follows we use the following notation: single and double primes denote
quantities measured in their respective comoving frames, while
subscripts 1 and 2 refer to quantities of the first and second components.

The energy density of synchrotron photons of the two components as measured in their comoving frames is given by
\eqb
u'_{\rm syn, 1} & = & \frac{\nu_p L^{\rm obs}_{\rm syn}(\nu_p) \large|_{1} }{4\pi cR_1^2 \delta_1^4} \\
u''_{\rm syn, 2} & = & \frac{\nu_p L^{\rm obs}_{\rm syn}(\nu_p)\large|_{2} }{4\pi cR_2^2 \delta_2^4},
\label{eq1}
\eqe
where we approximated the total synchrotron luminosity by its value at the peak frequency $\nu_p$ in each case. By
looking at the SEDs in Fig.~\ref{sed-ssc2} one sees that the error introduced by this assumption is small.
Taking also into account that $\nu_p L^{\rm obs}_{\rm syn}(\nu_p) \large|_{1} / \nu_p L^{\rm obs}_{\rm syn}(\nu_p) \large|_{2} \sim 0.6$
and using the values in Table~1 for the sizes and Doppler factors we find
that 
\eqb
\frac{u'_{\rm syn, 1}}{u''_{\rm syn, 2}} \simeq 1.7 \times 10^3.
\eqe
Because of this large difference in the energy densities of synchrotron photons, one can pose the question:
{\sl Can the energy density of synchrotron photons from the first component as seen in the
rest frame of the second one, be more important than the energy density of the internally produced?}

To answer, one has to calculate the quantity $u''_{\rm syn,1}$. 
We assume that the velocity vectors of the two components are parallel. Then
their relative velocity and Lorentz factor
are given by 
\eqb
\beta_{rel}&  = & \frac{\beta_2-\beta_1}{1-\beta_1 \beta_2} \\
\Gamma_{rel} & = & \Gamma_1 \Gamma_2 \left(1-\beta_1 \beta_2 \right).
\eqe
Assuming that $\Gamma_1 \approx \delta_1=18$ and $\Gamma_2 \approx \delta_2 =34$ we find
that $\beta_{rel}=0.56$ and $\Gamma_{rel}=1.2$. 

Using the invariance of $u(\epsilon, \mu)/\epsilon^3$ and the 
transformation of the solid angle 
\eqb
d\Omega''& = & \frac{2\pi}{\Gamma_{rel}^{2} \left(1-\beta_{rel} \mu'\right)^{2}} d\mu' 
\eqe
we find that 
\eqb
u''_{\rm syn,1} & = & \int d\epsilon'' \int d\mu'' u''_{\rm syn,1}(\epsilon'', \mu'') = \nonumber \\
& = & \int d\epsilon' \int d\mu' \Gamma_{rel}^2\left(1-\beta_{rel} \mu'\right)^2 u'_{\rm syn,1}(\epsilon', \mu') = \nonumber \\
& = & \frac{u'_{\rm syn,1}}{2}\Gamma_{rel}^2\int_{\mu{12}}^{1}d\mu' \left(1-\beta_{rel} \mu'\right)^2
\eqe
where $\mu_{12} = r_{12} / \sqrt{r_{12}^2+R_1^2}$. In the above we assumed an isotropic synchrotron photon field in the comoving
frame of the first component, i.e. $u'_{\rm syn,1}(\mu')=u'_{\rm syn,1}/2$. 
The result of the integration is a function of $r_{12}$ and $\beta_{rel}$:
\eqb
g(r_{12}, \beta_{rel}) = 1-\mu_{12} - \beta_{rel}(1-\mu_{12}^2)+\frac{\beta_{rel}^2}{3}(1-\mu_{12}^3).
\eqe
Combining the above, the condition $u''_{\rm syn, 1} \lesssim u''_{\rm syn, 2}/3$ is written as
\eqb
g(r_{12}, \beta_{rel}) < \frac{2 u''_{\rm syn,2}}{3 u'_{\rm syn,1} \Gamma_{rel}^2}.
\eqe
This results in  $r_{12} > 6\times 10^{16}$~cm. As long as the separation between the two is larger than
the radius of the larger emitting region, it is safe to neglect the synchrotron photon field of the first component.
For the opposite case, i.e.  what is the contribution of the synchrotron photon field of the second component
to the emission of the first one, a similar calculation is not necessary because the second component (i)  moves away from 
the first one and (ii)  it has a lower energy density ($u''_{\rm syn,2} << u'_{\rm syn,1}$).

\bibliographystyle{aa} 
\bibliography{pks}
\end{document}